\documentstyle[twocolumn,prl,aps,epsf,psfig]{revtex}
\begin{document}
 \draft

\title{Influence of quantum confinement on the ferromagnetism of (Ga,Mn)As
diluted
magnetic semiconductor}
\author{Sameer~Sapra,$^1$ D.D.~Sarma,$^{1,}$\cite{jnc} S.~Sanvito$^2$ and N.A.~Hill $^2$}
\address{$^1$Solid State and Structural Chemistry Unit,
Indian Institute of Science, Bangalore - 560012, India}
\address{$^2$Materials Department, University of California,
Santa Barbara, California 93106}
\date{\today}
\maketitle

\begin{abstract}

We investigate the effect of quantum confinement on the ferromagnetism
of diluted magnetic semiconductor Ga$_{1-x}$Mn$_x$As using a combination
of tight-binding and density functional methods.
We observe strong majority-spin Mn $d$-As $p$ hybridization, as well as
half metallic behavior, down to sizes as small as 20~\AA\ in diameter.
Below this critical size, the doped holes are self-trapped by the Mn-sites,
signalling both valence and electronic transitions. Our results imply
that magnetically doped III-V nanoparticles will provide a medium for
manipulating the electronic structure of dilute magnetic semiconductors
while conserving the ferromagnetic properties and even enhancing it in
certain size regime.
\end{abstract}



Over the last few years there has been an explosively rapid increase of activity
in two frontier areas of semiconductor research: diluted magnetic semiconductors
(DMSs) and semiconductor quantum dots.
The renewed interest in diluted magnetic semiconductors~\cite{kasuya,furdyna}
was motivated by the discovery of ferromagnetism in In$_{1-x}$Mn$_x$As and
Ga$_{1-x}$Mn$_x$As~\cite{ohno1,ohno2}.
Such ferromagnetic semiconductors are enabling materials for the emerging
technology of {\it spintronics}~\cite{prinz}.
Here, in addition to the charge degree of freedom of conventional
electronics, the spin degree of freedom can also be used so that
signals can be stored and processed simultaneously on the same
device component~\cite{kasuya}.
In addition, DMSs have been used to inject spin into normal
semiconductors~\cite{ohno4,fiederling}, where enormous spin life-time
and coherence~\cite{aws} have been demonstrated.
These are two key ingredients for the physical realization of solid state
quantum computing based on the spin degree of freedom~\cite{vincenzo}.

The second frontier area, the study of semiconductor quantum dots,
exploits the fact that by reducing the size of a semiconducting material,
typically below the Bohr exciton radius, it is possible to tune its electronic
and optical properties~\cite{yoffe}.
The changes result from the confinement of the electron and hole
wavefunctions by the potential of the finite sized semiconductor
nanoparticle, thereby increasing their energies as the particle
size is decreased. This
{\it quantum size effect} can have spectacular and desirable effects on
the material properties that have already been utilized for various device
purposes.
The most obvious applications are in tailoring of the band gap for specific
applications, such as the tuning of the absorption or emission energy in
electronic, electro-optical, opto-electronic or purely optical devices.

In this paper we explore the integration of the fields of diluted
magnetic semiconductors and
semiconductor quantum dots in a theoretical study of nanoparticles
made from the DMS material, (Ga,Mn)As.
Various experimental groups are actively pursuing the synthesis of magnetic
semiconducting nanoparticles, and the production of magnetically-doped
II-VIs, including CdS~\cite{cds}, ZnS~\cite{zns}, CdSe~\cite{cdse} and
ZnSe~\cite{znse}, has already been achieved.
The chemistry of III-V nanocrystal growth makes the synthesis
of (Ga,Mn)As nanoparticles more challenging, therefore a preliminary
theoretical investigation is sensible before launching a large synthetic
effort.
The main question that we answer in this work is whether the conditions
that lead to ferromagnetism in bulk (Ga,Mn)As - namely the combination of
strong Mn $d$-As $p$ hybridization in the majority spin band,
and a half-metallic band structure allowing hole-doping of the
majority spin channel - persist as the size of the system is reduced.
Our results clearly establish that quantum confinement does not have
a detrimental effect on the magnetic properties of (Ga,Mn)As nanoparticles
down to very small sizes. As a consequence it is possible to exploit the
quantum confinement effect to tune the
electronic properties of (Ga,Mn)As nanocrystals over a very wide size
range while retaining the  desirable ferromagnetism or even enhancing it.
Additionally, we see drastic changes in the electronic and
magnetic properties at very small ($<$~20~\AA\ diameter)
sizes, arising from
competition between different subparts of the electronic degrees of freedom.
These changes eventually lead to the destruction of the essential
electronic structural features required for ferromagnetism in
ultra-small nanoparticles.

There is in fact no {\it a-priori} reason that the magnetic properties
of
magnetically doped semiconductor nanoparticles should be the same
as those of the corresponding bulk system. Bulk (Ga,Mn)As is a
half metallic ferromagnet~\cite{us} (HMFM) with 100\%
spin polarization of the states at the Fermi energy, $E_F$.
Half-metallic ferromagnetism requires a very
specific arrangement of the various bands (in this case majority and minority
spin Mn 3$d$ with respect to the Ga and As
$s$, $p$ states) such that a gap opens for one spin channel only.
However
the energy levels of these states shift in nanoparticles due to the quantum
size effect, and the extent of the shift depends on both the size of the
nanoparticle and the effective masses of the electrons and holes.
In particular, the transition metal 3$d$ bands in a doped semiconductor are
rather flat compared to the $s$ and $p$ valence and conduction bands of the
host. This results in a rapid shift in the $s$ and $p$ derived
bands and a relative insensitivity of the $d$ band energy position as the
nanoparticle size is reduced. Thus
the underlying electronic structure of the doped nanoparticle will change
rapidly and may become incompatible with the HMFM state. Such a loss of
half-metallicity would in turn destabilize the ferromagnetism, limiting
the tunability of the electronic properties with size if the magnetic
properties need to be retained intact.

We calculate the electronic structure of Mn-doped GaAs nanoparticles as a
function of size using a second nearest neighbor tight-binding (TB)
method~\cite{SK}. We obtain the TB parameters by fitting to our density
functional band structure calculations of the corresponding bulk systems.
Our numerical DFT implementation uses a pseudoatomic orbital
basis set and pseudopotentials, within the local spin density
approximation (LSDA). Details of the method and its optimization can
be found in reference \cite{us}. The first step in the TB parameterization is
to fix the Ga and As on-site energies and the Ga-Ga, As-As and Ga-As hopping
integrals, by fitting to the GaAs band structure along several directions in
$k$-space. In this fit we use 35 $k$-points and we consider all the eigenvalues
up to the conduction band.
Then we repeat the fitting procedure for both the spin bands of MnAs. In this
case we do
not change the As-As hopping integrals and we allow the As on-site energies to
be only
rigidly shifted with respect to their values in GaAs. We do not allow
spin-splitting of the Mn $s$ and Mn $p$ orbitals, since only the $d$ orbitals
are magnetically
active. Finally we fix the remaining Ga-Mn parameters by fitting to the
calculated
band structure of a monolayer GaAs/MnAs superlattice.

Ga$_{1-x}$Mn$_x$As nanoparticles are generated by starting
with a central Ga atom surrounded by four As atoms, then
progressively adding successive shells of Ga and As atoms in the
bulk GaAs zincblende structure. Mn atoms are randomly substituted
at the Ga sites keeping the composition close to $x$~=~0.05 in every
case. The surface states of the clusters are passivated to quench
the dangling bonds. We have investigated particles ranging in size
from 6.0~\AA\ diameter (containing 17
atoms) to 71.0~\AA\ diameter (9527
atoms). We use exact diagonalization of the TB Hamiltonian for
the small clusters (up to 25.0~\AA\ containing 525 atoms) and the
Lanczos method for larger clusters.

In Fig. 1a, we show our calculated DFT majority and minority spin
densities of states (DOSs) of bulk Ga$_{1-x}$Mn$_x$As
with $x$~=~0.0625,
along with the partial Ga $s$, $p$, As
$p$ and Mn $d$ DOS. As previously reported~\cite{us}, the material
is a half metallic ferromagnet, with a gap in the minority spin
DOS at $E_F$, but no gap in the majority
channel~\cite{GapComment}. The metallic majority spin DOS at $E_F$
is comprised largely of As $p$ states, suggesting that hole doping
at the top of the valence band by the Mn$^{2+}$ 3$d^5$ ions, and
subsequent polarization of the mobile charge carriers by
interaction with the exchange-split energy levels of Mn $d^5$
\cite{sarmaprl}, is responsible for the HMFM state. The orbital
resolved DOS for the Mn 3$d$ states is qualitatively different for
majority and minority spins. The majority channel has extensively
mixed states over the -1 to -4~eV energy range due to the strong
hybridization between the Mn 3$d$ and the As $p$ states.
In contrast, the DOS of the Mn 3$d$ states for the minority spin
shows a sharp feature at $\sim$1.5~eV above $E_F$ with minimal
admixture from other states. This difference is due to the fact that
in the majority spin the Mn 3$d$ states are in the middle of the valence
band of GaAs, thereby mixing extensively with the As $p$.
In contrast, the minority spin Mn 3$d$ states fall in the GaAs gap,
thereby remaining relatively unmixed. The DOS for the extended solid
obtained from the tight binding Hamiltonian is shown in Fig. 1b.
All the features of the DOS, including the HMFM behavior, are
reproduced extremely well by the TB calculation, confirming the
accuracy of our parameterization.

In the limit of large size, the electronic properties of
semiconductor nanoparticles must approach asymptotically those of
the corresponding bulk semiconductor. The calculated DOS for a
71.0~\AA\ diameter cluster, shown in Fig.
1c, is indeed similar to that of the infinite solid.
Clearly smaller particles are required to see any strong effect of
quantum confinement. Our calculations of the electronic structures
of progressively smaller-sized nanoparticles (not shown here) indicate
that qualitatively similar electronic properties persist down to a
very small size ($\sim$25~\AA\ diameter) although the band gap
increases monotonically with decreasing size as expected. In Fig.
2a we show the calculated majority and minority spin DOS
for a 25.0~\AA\ diameter nanoparticle. We see that even such a
small particle is both half-metallic and
has strong Mn $3d$ - As $p$ hybridization, therefore has all the
indications for being ferromagnetic. This is remarkable, since it
implies that it is possible to exploit the tunability of
nanoparticles to tailor-make materials with specific electronic
properties, without sacrificing the ferromagnetism. For example,
the gap (1.4~eV) in the down-spin channel in this cluster is
substantially larger than that in the larger 71.0~\AA\ diameter
particle (0.55~eV), or in the bulk (0.4~eV). So, for example, such
nanoparticles could be used as a source of polarized light over a
broad range of wavelengths.

However the 25.0~\AA\ diameter (Ga,Mn)As nanoparticle is
close to the size limit below which the
essential properties of the bulk magnetic semiconductor
are no longer retained. At smaller sizes, we see
remarkably rapid changes in the electronic structures with decreasing
size, signalling a complete change of electronic and magnetic
properties. In fact we observe a cross-over behavior, akin to a
valence as well as a
metal-insulator transition, that is driven by the quantum
size effect.
We illustrate this in Fig. 2 where we also plot our calculated
DOSs of 18.5 and 12.1~\AA\ diameter nanoparticles. Compared to
the 25.0~\AA\ nanoparticle, the DOS of
the 18.5~\AA\ particle (Fig. 2b) shows
a substantial
increase of the energy gap in the down-spin channel,
as a consequence of the enhanced confinement, and
a significant
reorganization of the orbital contributions to the DOS near $E_F$,
particularly in the majority channel. We find that the states close
to $E_F$ develop a significant amount of Mn $d$ character,
whereas the larger particles (Fig. 1c)
and the infinite solid (Figs. 1a and 1b) have almost negligible
contribution from the Mn $d$ states at $E_F$. Simultaneously, the
Mn $d$ states in the majority channel are reduced in bandwidth
compared to the larger clusters. For example the
majority spin Mn $d$ band width in 12.1~\AA\ diameter
nanoparticles (shown in Fig. 2d) is $\sim$1~eV
narrower than that in the bulk solid. These results are easily
understood in terms of quantum confinement effects: the GaAs $s$ and
$p$ states, with low effective masses, shift very rapidly compared
to the higher effective mass Mn $d$ states. As the GaAs valence and
conduction bands shift away from each other,
majority spin Mn $d$ states begin to appear at the top
of the valence band. There is a decrease of the As $p$
band width upon size reduction which is also primarily
responsible for the reduced bandwidth of the Mn $d$ band in the
majority channel, through the strong hybridization between
the Mn $d$ and the As $p$ orbitals.
In contrast, the Mn $d$ minority spin states continue to
form a narrow peaked structure a couple of eVs above the
Fermi energy even for the smallest nanoparticles.

There are interesting consequences of these movements of
the energy levels, {\it via} changes in the Mn $d$
characters in the highest occupied
molecular orbital (HOMO) with the size of the clusters
(Fig.~3). The percentage Mn $d$ character at $E_F$ is $\sim$8\%
in the extended solid as well as in the large clusters, indicated
by the horizontal dotted line in Fig. 3. The Mn $d$ character
begins to increase appreciably below about 35~\AA\ due to the
downward shift of the valence band. This approach of the Fermi energy
towards the Mn $d$ majority state will in fact enhance the effective
exchange coupling between the localized Mn $d^5$ spins and the
conduction band \cite{sarmaprl} and thereby enhance the $T_c$
with decreasing cluster size. However, as the shift of the top
of the valence band continues beyond the Mn $d$ majority state,
the holes that were originally doped at the top of the valence
band are increasingly trapped by the Mn sites, driving a Mn valence
transition. This is signalled by the sharp rise in the Mn character
in HOMO (Fig. 3) below $\sim$20 \AA\ cluster size with as much as 92\%
Mn $d$ contribution for the 6 \AA\ clusters. This is also clearly shown
by the change in the average Mn $d$ occupancy, $<n_d>$. It
should be noticed that this quantity cannot be used to determine
directly a formal oxidation state of the Mn sites, due to the
presence of substantial covalent admixture of Mn states with the
GaAs states. We calculate that $<n_d>$ is about
5.23 in the 25.0~\AA\ nanoparticle, corresponding to the formally
divalent Mn$^{2+}$ state. However, $<n_d>$
changes very rapidly with decreasing size below 25.0~\AA\ to
become 4.37 for the smallest sized 6.0~\AA\ cluster. This change
of almost unity in $<n_d>$ between 25.0~\AA\ and 6.0~\AA\ clusters
indicates a valence transition with the Mn$^{2+}$ ions
self-trapping the doped holes, resulting in Mn$^{3+}$ and the
removal of the doped holes from the top of the valence band of
GaAs. This, along with the absence of any substantial kinetic stabilzation
{\it via} the strong antiferromagnetic exchange splitting of the
conduction states \cite{sarmaprl} due to
the placement of the valence band below the Mn majority
spin state, will
occur in particles smaller than $\sim$20~\AA\ diameter,
leading to an insulating state which is unable to sustain
ferromagnetism.

In conclusion we have studied the effect of quantum size confinement
on the magnetic properties of (Ga,Mn)As nanoparticles, using an LSDA-derived
TB parameterization.
We find that the nanoparticles retain the desirable ferromagnetic
properties observed in bulk (Ga,Mn)As down to diameters as small as
$\sim$20~\AA\, even enhancing the magnetism over a certain size regime.
For particles larger than this critical size, we find
an half-metallic density of states, with a polarized As $p$ hole at the Fermi
energy. Smaller clusters show a large contribution of the Mn $d$ state at $E_F$,
eventually leading to the disappearance of the magnetic state.
These results suggest that the electronic properties (in particular the band
gap and also possibly $T_c$) of diluted magnetic semiconductor nanoparticles can be manipulated
by controlling the cluster size very effectively.

S.Sapra and D.D.S. acknowledge financial support from DST,
Government of India. S.Sanvito and N.A.H. acknowledge the ONR (N00014-00-10557),
NSF-DMR (9973076) and ACS PRF (33851-G5), for financial
support. Part of this work made use of MRL Central Facilities supported
by the National Science Foundation (DMR96-32716).


\begin{figure}[h]
\caption{\label{figdos} Density of states for Ga$_{1-x}$Mn$_x$As
with $x$~=~0.0625 calculated with different methods: (a) LSDA
calculation for the bulk, (b) tight-binding calculation for the
bulk, (c) tight-binding calculation for a 71.0~\AA\ cluster
containing 9527 atoms (the cluster DOS has been broadened by a
Gaussian with 0.3~eV FWHM).}
\end{figure}

\begin{figure}[h]
\caption{\label{figshells} Density of states and partial As $p$
density of states for three (Ga,Mn)As finite clusters: (a) 9
shells and 25.0~\AA\ diameter, (b) 7 shells and 18.5~\AA\
diameters, (c) 5 shells and 12.1~\AA\ diameter, and (d) partial
Mn $d$ density of states for these three clusters. In all the panels
the zero of energy corresponds to the $E_F$ of bulk Ga$_{1-x}$Mn$_x$As
with $x$~=~0.0625. Note that the
minority spin band gap increases with decreasing the size of the
nanocrystal. The majority spin band gap collapses for the
12.1~\AA\ nanocrystal. Also note that the DOS have been broadened
by a Gaussian with 0.3~eV FWHM. }
\end{figure}

\begin{figure}[h]
\caption{\label{figpercent} (a) Mn $d$ character of the HOMO as a
function of the cluster size.}
\end{figure}


\begin{thebibliography}{99}
\bibitem{GapComment}
We have made no attempt in this paper to correct for the well-known
underestimation of the band gap within the LSDA. However, although
the absolute values of our calculated band gaps are all too small,
the shifts with size, which are dominated by occupied valence band
states, should be reproduced quite accurately.
\end{thebibliography}
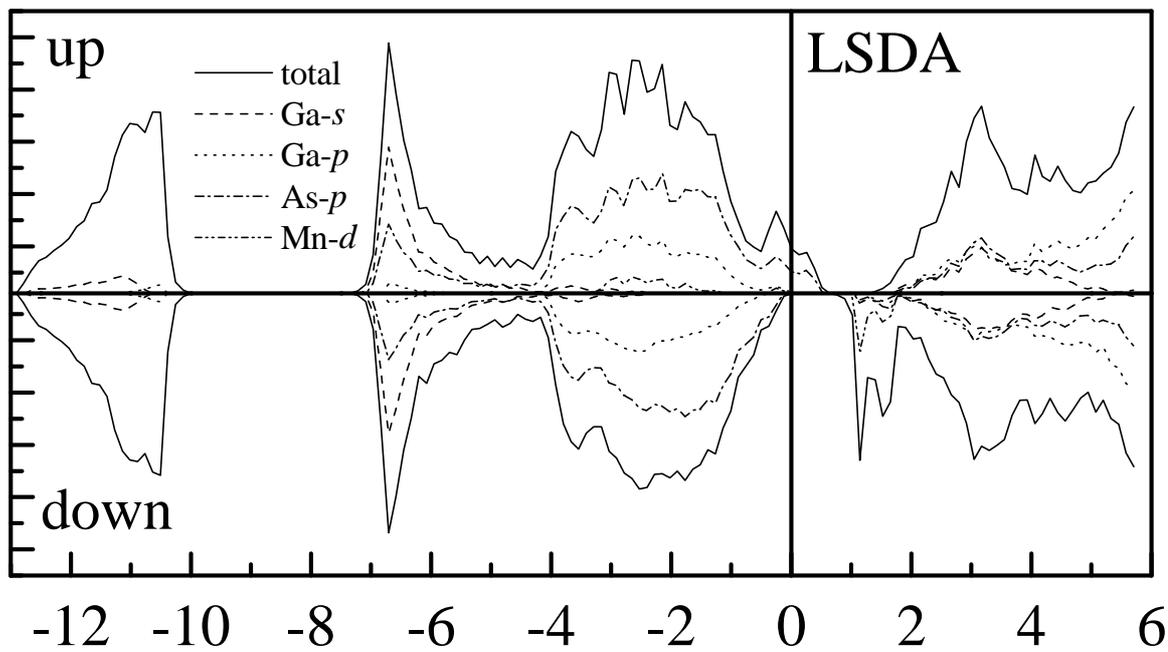
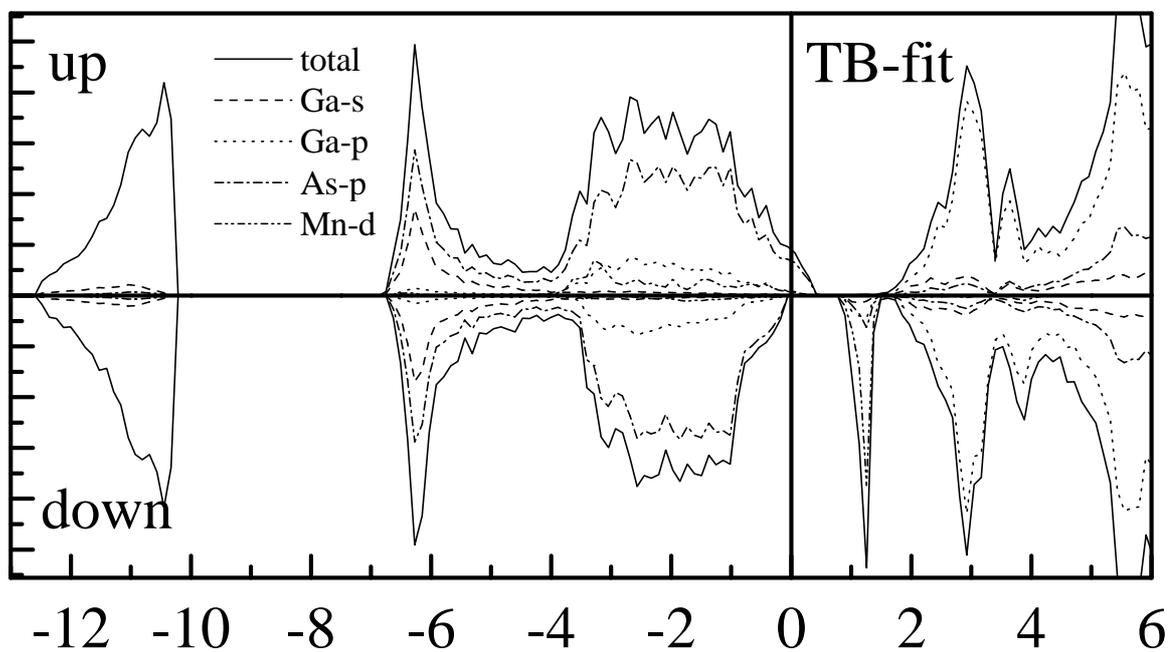
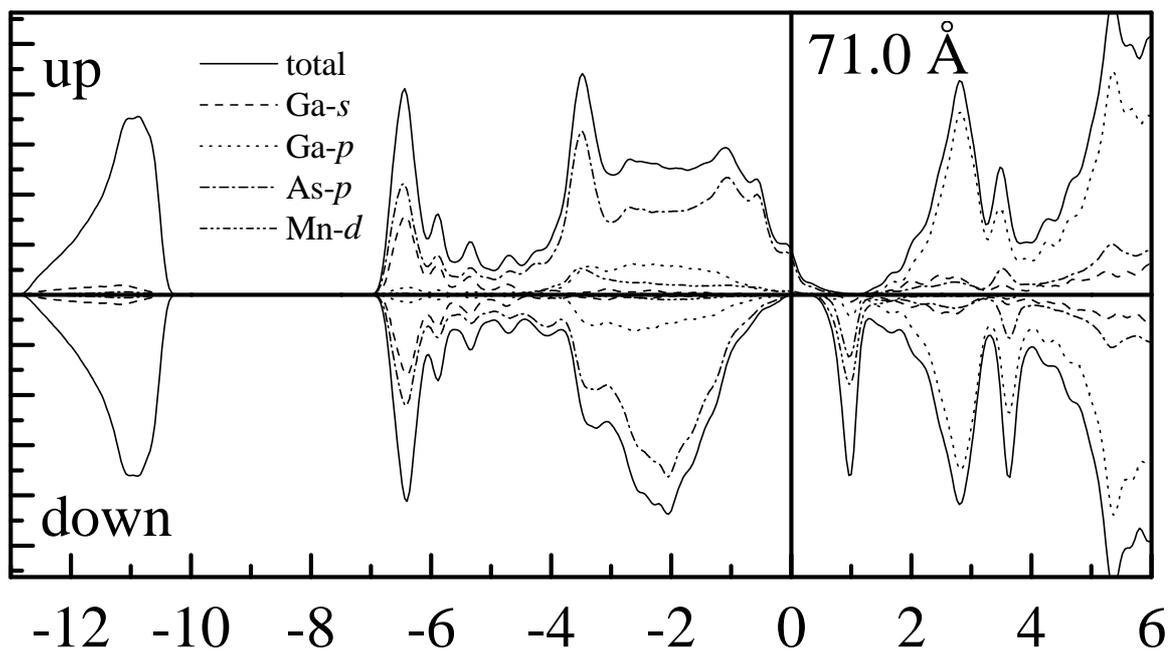
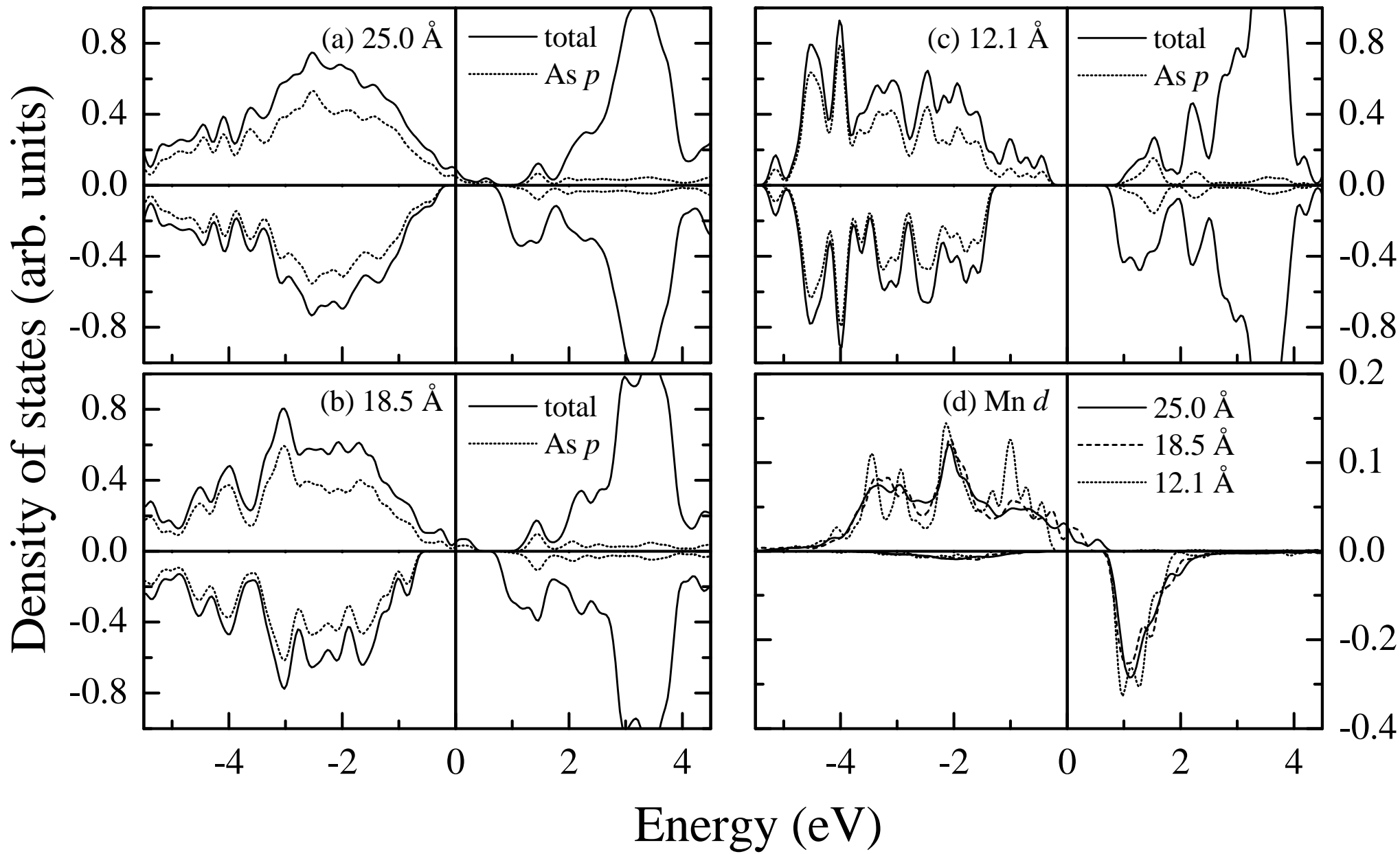
\end{document}